\documentstyle[eqsecnum,aps,preprint]{revtex}
\def\laq{\ \raise 0.4ex\hbox{$<$}\kern -0.8em\lower 0.62
ex\hbox{$\sim$}\ }
\def\gaq{\ \raise 0.4ex\hbox{$>$}\kern -0.7em\lower 0.62
ex\hbox{$\sim$}\ }
\def\half{\hbox{\magstep{-1}$\frac{1}{2}$}}

\def\NPB{{\em Nucl. Phys.} B}
\def\PLB{{\em Phys. Lett.}  B}
\def\PRL{{\em Phys. Rev. Lett. }}
\def\PRD{{\em Phys. Rev.} D}

\begin{document}

\preprint{\vbox{\baselineskip=12pt
\rightline{BGU-PH-98/05}
\vskip0.2truecm
\rightline{TAUP-2486-98}
\vskip1truecm}}

\vskip 2 cm

\tighten

\title{Dark Matter Axions in Models of String Cosmology}

\author{Ram Brustein} 
\address{Department of Physics,
Ben-Gurion University,
Beer-Sheva 84105, Israel\\
email: ramyb@bgumail.bgu.ac.il}

\author{Merav Hadad} 
\address{School of Physics and Astronomy,
Beverly and Raymond Sackler Faculty of Exact Sciences,\\
Tel Aviv University, Tel Aviv 69978, Israel\\
email: meravv@post.tau.ac.il}

\maketitle
\begin{abstract}
Axions are produced during a period of dilaton-driven inflation by amplification
of quantum fluctuations. We show that for some range of string cosmology
parameters and some range of axion masses,  primordial axions may constitute
a large fraction of the present energy density in the universe in the
form of cold dark matter. Due to the periodic nature of the axion
potential energy density fluctuations are strongly suppressed. The spectrum of
primordial axions is not thermal, allowing a small fraction of the axions to
remain relativistic until quite late. 
\end{abstract}


\section{Introduction}
Axions are hypothetical particles, invented to solve the strong CP problem 
\cite {pq,axions}. Many ongoing experimental efforts aim to detect
axions, but have yet to produce evidence for their existence. Cosmological and
astrophysical implications of axions are well studied \cite{eu,axrev}, in
particular, axions are among the leading candidates for providing the missing
dark mass in the universe.

String theory possesses many axion candidates
\cite{witten,ck}  of the ``invisible axion" type
\cite{invax}. We will focus on the ``model-independent axion" \cite{witten}
which respects a Peccei-Quinn symmetry to all orders in perturbation theory.
There are no good general arguments that QCD provides the dominant contribution
to the potential energy of any of the stringy axions, including the model
independent axion. However, there are some theoretical conditions under
which the model-independent axion could be the axion that
solves the strong CP problem. In any case, we will assume that this is indeed so.

We consider axion production in models of string cosmology which realize the
pre-big-bang scenario \cite{sfd,pbb}. In this scenario the evolution of
the universe starts from a state of very small curvature and coupling and
then undergoes a long phase of dilaton-driven kinetic inflation  and at some
later time joins smoothly standard  radiation dominated cosmological evolution,
thus giving rise to a singularity free inflationary cosmology.
Axions are produced during the
period of dilaton-driven inflation by the standard mechanism of amplification of
quantum fluctuations\cite{mukh}. 

The spectrum of relic axions depends on 
their potential and interactions, and on some of the string cosmology model
parameters.  By applying simple constraints, such as requiring that the energy
density of the universe does not exceed the critical density at different stages
of the evolution, we are able to constrain parameters of string cosmology models
and axion potential and find a consistent parameter range in which most of the 
energy of the universe today is in the form of cold dark matter axions whose
origin is quantum fluctuations from the pre-big-bang. This consistent range
overlaps with the range in which relic gravity wave background produced during
the dilaton-driven inflationary phase could be detected by planned gravity wave
experiments \cite{bggv}. The same parameter range could perhaps lead to
formation of observable primordial black holes \cite{cope1}.

\section{The Model}

We assume that the model independent axion receives the dominant part of its
potential from QCD instantons, therefore, roughly speaking, the axion is
massless until the universe cools down to a temperature $T_{QCD}\gaq
\Lambda_{QCD}$ at time  $t=t_{QCD}$. A more sophisticated
estimate as in \cite{eu,axrev} will be used later. The axion then develops a
periodic potential of overall approximate strength $\Lambda_{QCD}^4$ (recall that
$\Lambda_{QCD}\sim 200 MeV$), and period which is apriori a free parameter
$f_{PQ}$. The scale $f_{PQ}$ is typically less than $10^{16}GeV$ resulting in an
axion mass  $m_a \gaq 10^{-10}eV$. Astrophysical constraints further bound
$f_{PQ}$ from below $f_{PQ}\gaq 10^9 GeV$  resulting in an axion mass 
$m_a \laq 10^{-2}eV$. We will discuss cosmological constraints on $f_{PQ}$ in
more detail later. The explicit
form  of the axion potential that we will assume $V(\psi)=\half
V_0\left(1-\cos(\frac{\psi}{\psi_0})\right)$, depends on two parameters, 
$\psi_0$, related to the Peccei-Quinn scale $f_{PQ}$, and $V_0=m_a^2
\psi_0^2$. For QCD axions  $V_0\simeq f_{\pi}^2 m_\pi^2\sim \Lambda_{QCD}^4$.

The model of background evolution we adopt in this paper is a simplified
model used also in \cite{bh}. The evolution of the universe is divided into four
distinct  phases, the first phase is a long the dilaton-driven
inflationary phase, the second phase is a high-curvature string phase of
otherwise unknown properties, followed by ordinary Friedman-Robertson-Walker 
(FRW) radiation dominated (RD) evolution and then a standard FRW matter dominated
(MD) evolution.  We assume throughout an isotropic and homogeneous
four dimensional flat universe, described by a FRW metric. The model is
described in full detail in \cite{bh}, and we reproduce here only its important
features.  We note in particular that the axion field  is assumed to have a
trivial vacuum expectation value during the inflationary phase. 

The dilaton-driven inflationary phase lasts until time $t =t_s$, 
and while it lasts the scale factor $a(t)$ and the dilaton
$\phi(t)$ are given by the solution of the lowest order string-dilaton-gravity
equations of motion, the so-called $(+)$ branch vacuum. 
The string coupling parameter $e^\phi=g_{string}^2$, 
in the models that we consider is, of course,  time dependent.
Both curvature and coupling
$e^\phi$ are growing in this phase, which is expected to last until curvatures
reach the string scale  and the background solution starts to deviate
substantially from the lowest order solution. For ideas about how this
may come about see \cite{exit}.

The string phase lasts  while  $t_s<t <t_1$. We assume that
curvature stays high during the string phase. As in \cite{peak}, we assume
that  the string phase ends when curvature reaches the string scale $M_s$, 
$H(t_1)\simeq M_s$.  We
parametrized our ignorance about the string phase background, as in \cite{bggv},
by the ratios of the scale factor and the string coupling
$g(t)=e^{\phi(t)/2}$,  at the beginning and end of the string phase
$z_S=a_1/a_S$ and $g_1/g_S$, where $g_1=e^{\phi(t_1)/2}$ and 
$g_S=e^{\phi(t_S)/2}$, where $a_S=a(t_s)$ and $\phi_S=\phi(t_s)$.   
We take the parameters to be in a range we consider
reasonable. For example, $z_S$ could be in the range $1<z_S<e^{45}\sim 10^{20}$,
to allow a large part of the observed universe to originate in the
dilaton-driven phase, and $g_1/g_S>1$, assuming that the coupling continues to
increase during the string phase and $10^{-3}\laq g_1\laq 10^{-1}$ to agree with the
expected range of string mass (see e.g. \cite{peak}). Some other useful
quantities that we will need are $\omega_1$, the frequency today, corresponding
to the end of the string phase, estimated in \cite{peak} to be $\omega_1\sim
10^{10}Hz$, and the frequency $\omega_S= \omega_1/z_S$, the frequency today
corresponding to the end of the dilaton-driven phase.

Standard RD phase and then MD phase are assumed to follow the string  phase. 
The dilaton is taken to be strictly constant, frozen at its value today.  

We have presented our assumptions about background evolution
and axion potential in great detail, and will use them as presented, even though
many of the assumptions can be either relaxed (without affecting dramatically our
results), or improved to take into account additional known effects. However,
each change adds an additional level of complication by adding parameters and
assumptions,
 and we preferred to keep the discussion as simple as possible to capture the
essential physics. Nevertheless, we do mention from time to time a possible
alternative or generalization.

\section{Primordial Axions}

The spectrum of axionic perturbations produced during the dilaton-driven 
inflation is approximately given by \cite{cope2} (see also \cite{bh,bmuv})
\begin{equation}
\psi_k = {\cal N}_k a_S^{-1} g_S^{-1} k_S^{-1/2} 
\left(\frac{k}{k_S}\right)^{-\sqrt{3}}, k<k_S,
\end{equation}
where ${\cal N}_k=2^{\sqrt{3}-1} \Gamma(\sqrt{3})/\sqrt{\pi}$ and $k_S=\omega_S
a(t)$. The r.m.s amplitude of the perturbation in a logarithmic $k$ interval
is defined in a  standard way $\delta\psi_k\equiv k^{3/2} \psi_k$. Using the
relation  $k_S/a_S=k_1/a_1=\omega_1(t_1)$, the assumption
$\omega_1(t_1)=M_s$ and  the relation between the string mass and the Planck mass
in weakly coupled string theory, $M_s= M_p g_1$ we obtain 
\begin{equation} 
\delta\psi_k = {\cal N}_k M_p\ \frac{g_1}{g_S}  
\left(\frac{k}{k_S}\right)^{3/2-\sqrt{3}}, 
\label{delpsi}
\end{equation}
for perturbations outside the horizon, and 
\begin{equation} 
\delta\psi_k = {\cal N}_k M_p\ \frac{g_1}{g_S}  
\left(\frac{k}{k_S}\right)^{1/2-\sqrt{3}}
\left(\frac{H(t)}{\omega_S(t)}\right), 
\label{delpsiin}
\end{equation}
for perturbations that have reentered the horizon during RD, before the
axion potential is generated.  

The ratio of energy density in axions per logarithmic
frequency interval  to the critical density $\frac{d\Omega_a}{d\ln\omega}$, for
perturbations that reenter the horizon during RD, before the axion potential is
generated, is given by \cite{bh} \begin{equation}
\frac{d\Omega_a}{d\ln\omega}={\cal C}
g_1^2\left(\frac{g_1}{g_S}\right)^2
\left(\frac{\omega}{\omega_S}\right)^{2\sqrt{3}-3},\hspace{.3in} \omega<\omega_S,
\label{Omegaax}
 \end{equation}
where ${\cal C}$ is a numerical factor 
which we will ignore in the following. Note that the spectral index 
$3-2\sqrt{3}\simeq -0.46$ is negative, and therefore most of the energy is
contained in the low-frequency modes.

The total energy density within the horizon, at a given time, is dominated by the
lowest frequency which is just reentering the horizon,
\begin{equation}
\Omega_a(t)=\int\limits_{H(t)}^{\omega_S(t)}
\frac{d\Omega_a}{d\ln\omega} d\ln\omega \simeq g_1^2
\left(\frac{g_1}{g_S}\right)^2
\left(\frac{H(t)}{\omega_S(t)}\right)^{3-2\sqrt{3}}, 
 \end{equation}
where
$\omega_S(t)=\omega_1(t)/z_S$. If $H(t)>\omega_S(t)$ then the
total energy density in axions produced during dilaton-driven phase simply
vanishes (We will discuss an estimate for the axions produced during the string
phase later on). 
Since $H(t)\propto T^2(t)$ and $\omega_S(t)\propto T(t)$, and since $H(t_1)=M_s$, then
$\frac{\omega_S(t)}{H(t)}= T_1/(T z_S)$, and therefore 
\begin{equation}
\Omega_a(t)\simeq g_1^2\left(\frac{g_1}{g_S}\right)^2
\left(\frac{M_s}{T(t) z_S}\right)^{2\sqrt{3}-3}.
\label{Omegafinal}
 \end{equation}

To ensure standard RD cosmology at late times  we must require that
the energy density in axions remains smaller than critical $\Omega_a<1$. It is
enough to require this at the lowest temperature possible, i.e., at the
temperature just as the axion potential is generated,    
$\Omega_a(t_{QCD})\simeq g_1^2\left(\frac{g_1}{g_S}\right)^2
\left(\frac{M_s}{T_{QCD} z_S}\right)^{2\sqrt{3}-3}<1$. Using $M_s=g_1 M_p$,
we obtain the following condition 
\begin{equation}
z_S > \frac{g_1 M_p}{T_{QCD}}\left[ \left(\frac{g_1}{g_S}\right)^2
g_1^2 \right]^{1/(2\sqrt{3}-3)}.
\label{cond} 
\end{equation}
Note that since $\frac{M_p}{T_{QCD}}\sim 10^{19}$, unless $z_S$ is large enough
condition (\ref{cond}) can be satisfied only if $g_1$ is unacceptably small.
There is, however, a reasonable range of parameters for which condition 
(\ref{cond}) is indeed satisfied, for example, if  $g_1=10^{-3}, g_S/g_1=1/10$,
then (\ref{cond}) implies $z_S\gaq 3\times 10^7$ and if $g_S\simeq g_1\simeq 0.01$ then
$z_S\gaq 3\times 10^{8}$. If $g_1\sim g_S$, condition (\ref{cond})
simplifies to $z_S\gaq g_1^5 M_p/T_{QCD}$. When (\ref{cond}) is saturated, axions
provide near closure density of the universe just before the axion potential is
generated. Condition (\ref{cond}) is valid for standard adiabatic RD evolution.
If some intermediate period of matter domination or entropy production is
assumed, condition (\ref{cond}) is relaxed. 

So far we have considered only axions that were produced during the
dilaton-driven phase and ignored axions that were produced during the
subsequent string phase. We would like to show that it is reasonable to
neglect axion production during the string phase by giving an estimate based on
the extrapolation used in \cite{bmuv},  which assumes constant $H$ and 
$\dot\phi$ during the string phase. The resulting energy density 
$\Omega_a^{sp}$ is given by 
\begin{equation}
\frac{d\Omega_a^{sp}}{d\ln\omega}\simeq
g_1^2\left(\frac{g_1}{g_S}\right)^2
\left(\frac{\omega}{\omega_S}\right)^{-2\zeta},\hspace{.3in}
\omega_S<\omega<\omega_1, 
\label{Omegasp}
 \end{equation}
where the spectral index $\zeta=\ln (g_1/g_S) / \ln z_S$, is
positive and therefore the energy density decreases with frequency. The
total additional energy in axions produced during the string phase
$\int\limits_{\omega_S(t)}^{\omega_1(t)}
 \frac{d\Omega_a^{sp}}{d\ln\omega}d\ln\omega$, is up to a numerical factor
$\sim g_1^2\left(\frac{g_1}{g_S}\right)^2 \frac{1}{2\zeta}$,
which is indeed negligible (for large $z_S$ and reasonable $g_1$, $g_S$)
compared with the energy density (\ref{Omegafinal}) in axions produced during the
dilaton-driven  phase. The same conclusion is expected as long as the spectrum
of axions produced during the string phase continues to decrease.

 If the model-independent axion were to remain massless its total energy
just before matter radiation equality would be given by  
$
\Omega_a(t_{eq})\simeq g_1^2\left(\frac{g_1}{g_S}\right)^2
\left(\frac{M_s}{T_{eq} z_S}\right)^{2\sqrt{3}-3}, 
$ (recall that $T_{eq}\sim 1 eV$). Axions would overclose the universe and lead
to an unacceptable cosmology, unless the parameters of string cosmology, and
in particular $z_S$ are pushed to uncomfortable values.

\section{Dark Matter Axions}

We turn to discuss the effects of the axion potential as it turns on when
the universe  cools down to QCD temperatures. If we try to approximate the axion
potential by a quadratic potential, a common practice in most investigations, we
encounter a puzzle. The axion energy density becomes formally divergent as soon
as the axion potential turns on! (if we assume that the dialton-driven phase
lasted only a finite time then the formal divergence is replaced by a singular
dependence on the duration of the dilaton-driven phase).  The relative energy
density in axions, assuming a quadratic potential 
$\frac{d\Omega_a^{QP}}{d\ln\omega}$, was computed in \cite{bh} and we reproduce
here its low frequency part,
 \begin{equation}
\frac{d\Omega_a^{QP}}{d\ln\omega}\simeq
g_1^2\left(\frac{g_1}{g_S}\right)^2 \frac{\sqrt{M_s m_a}}{\omega_1}
\left(\frac{\omega}{\omega_S}\right)^{3-2\sqrt{3}},\hspace{.3in}
\omega<\omega_m, \omega_S. 
\label{Omegam}
 \end{equation}
Once the potential is generated, all the low frequencies
reenter the horizon at once, so to obtain the total energy density inside the
horizon we need to integrate $\frac{d\Omega_a^{QP}}{d\ln\omega}$ from the
minimal amplified frequency $\omega_{min}$, which is either zero, if the
duration of the dilaton-driven phase is infinite, or exponentially small if the
duration is finite but large, 
$\Omega_a^{QP}(t)=\int\limits_{\omega_{min}}
\frac{d\Omega_a^{QP}}{d\ln\omega} d\ln\omega$. The lower frequency part of the spectrum
yields a divergent contribution, proportional to 
${\omega_{min}}^{3-2\sqrt{3}}$ (recall $3-2\sqrt{3}\simeq -0.46$). This
result does not make sense.

The resolution of the puzzle depends crucially on the periodic nature of
axion potential $V(\psi)=\half V_0\left(1-\cos(\frac{\psi}{\psi_0})\right)$. 
This point was first understood by Kofman and Linde \cite{koflinde}, 
and we have adopted their ideas to our particular situation. 
First, the total  potential energy is limited to $V_0$ and does not continue to increase
indefinitely as the axion field increases, providing a ``topological cutoff" on
the total axionic energy density and as important, large fluctuations in the
axion field are also  ``topologically cutoff", producing exponentially small
energy density perturbations.  Large fluctuations lead to a uniform
distribution of the axion field inside the horizon, with very small statistical
fluctuations.

The axion potential is highly non-linear, therefore it is not possible
to solve the perturbation equation mode-by-mode. In \cite{koflinde}, 
the following strategy is suggested. Consider the axion field
$\psi(\vec{r},t)$ at the time when the axion potential is turned on. The low $k$
Fourier modes $\psi_k, k/a(t_{QCD})<H(t_{QCD})$, 
provide an essentially constant field $\psi_c$ 
across the horizon. The value of $\psi_c$ is random, and in our case it is
determined statistically by a Gaussian distribution $P(\psi_c)$ with zero average
and standard deviation 
$\sigma_c=\sqrt{\int\limits_{k_{min}}^{aH} d\ln k|\delta\psi_k|^2}$. Since in
all cases that we will be interested in, $\psi_0< 10^{16} GeV$, and 
$\delta\psi_k>M_p$ for all $k<aH$ (see eq.(\ref{delpsi})), 
the width $\sigma_c$
is much larger than the period of the axion potential $\sigma_c \gg
2\pi\psi_0$.  The constant value $\psi_c$ becomes essentially uniformly
distributed among all possible values. 
The  average energy density in the non-relativistic part of the axion
field $\rho_a=\langle  V\left(\psi(\vec r)\right) \rangle$ 
is given by  
\begin{equation}
\rho_a=\int \half V_0\left(1-\cos\left(\frac{\psi}{\psi_0}\right)\right)
P(\psi) d\psi \simeq \half V_0,  
\label{rhoa}
\end{equation}
with exponentially small corrections. Note that eq.(\ref{rhoa}) is valid for all
reasonable values of $z_S$ and $g_S$, $g_1$. Regardless of the fraction of
relativistic axions which exists at $t_{QCD}$, the low momentum modes with
wavelength larger than the horizon contribute a constant energy density.
The procedure that we outlined above can be repeated for any scale $\ell$, 
separating modes of $\psi(\vec r, t)=\psi_c(\ell^{-1})+\widetilde\psi$, where 
$\widetilde\psi$ contains only modes with $k<\ell^{-1}$.

The constant axion field $\psi_c$ starts to coherently oscillate around the nearest
minimum of the potential.
Using  completely standard arguments \cite{eu,axrev,cosmaxion}, we may obtain a
bound on $m_a$  (or equivalently on $\psi_0$) by requiring that the energy
density  in the coherent axion oscillations  be subcritical at the beginning of
MD epoch. This requirement leads to the  standard bounds on the axion mass,
except the possibility that the axion ``starts" at a special point seems less
viable. We may evaluate the number of axion particles at the initial
time when the potential is turned on ( defined by the condition $m(T_{QCD})=3
H(T_{QCD})$), $n_a=\rho_a/m_a$. Using $m_a(T)^2=V_0(T)/\psi_0^2$ we may estimate
$V(T_{QCD})\sim T_{QCD}^4 \psi_0^2/M_p^2$, leading to  the standard
estimate
$
\Omega_a h^2 \sim \frac{10^{-6}eV}{m_a},
$
where  $h$ is todays Hubble parameter in units of 100 km/Mpc/sec.
Requiring subcritical $\Omega_a$ leads to the standard bound on $\psi_0$,
$\psi_0\laq 10^{12} GeV$ and $m_a\gaq10^{-6} eV$.

In string theory, natural values of $\psi_0$ are approximately $10^{16}
GeV$, which, if taken at face value, would lead to overclosure of the universe
with axions many times over. Two possible resolutions have been suggested 
\cite{banksdine1,banksdine2,ch} 
to allow our universe to reach its old age of today. First, that somehow,
perhaps involving some strong coupling string dynamics, the low energy effective 
$\psi_0$ is some orders of magnitude below  $10^{16} GeV$, and the second is
that some non-standard matter domination epoch, or some late entropy production,
has occurred in between $T_{QCD}$ and  nucleosynthsis epoch. 
Our results cannot shed further light on this
problem, but they do reinforce the need for a resolution. If the resolution of
the $\psi_0$ problem requires strongly coupled string theory $g_1> 1$, some of
our assumptions should be changed but most likely our estimates are still valid,
and therefore our results are probably qualitatively correct also in that case. 
Of course, another possible resolution is that the model independent axion is not
the QCD axion.

We turn now to the question of energy fluctuations. Since there are large
fluctuations  in the axion field (\ref{delpsi}), (\ref{delpsiin}), we should worry about large
energy fluctuations which will cause  unacceptable deviations from isotropy and
homogeneity, affecting either  nucleosynthsis or the cosmic
microwave background. However, as explained in \cite{koflinde}, 
these perturbations are
suppressed. Fluctuations in the axion energy density at a scale $\ell\sim
k^{-1}$ can be computed by using the relation
\begin{equation}
\int\limits_{0}^{\infty} \left(\delta\rho_a^2\right)_k \frac{\sin{k r}}{kr}
d\ln k= \left\langle V\left(\psi(\vec x)\right) 
V\left(\psi(\vec x+\vec r)\right) \right\rangle- 
\left\langle V\left(\psi(\vec x)\right)\right\rangle^2
\label{delrho}
\end{equation}
 In previous expressions 
$\left\langle\cdots \right\rangle$ denotes
either vacuum expectation values of operators or statistical averages.

To evaluate   (\ref{delrho}) we need   
\begin{equation}
\left\langle \cos\left(\psi(0)\right)
\cos\left(\psi(\vec{r})\right) \right\rangle 
-\left\langle \cos\left(\psi(0)\right) \right\rangle^2 =
e^{- \left\langle \psi^2(0) \right\rangle }
 \left[ 
{\rm Cosh} \left( \left\langle \psi(0)\psi(\vec r)\right\rangle\right)-1
\right], 
\end{equation}
using it we obtain  (for the case
$\langle\psi\rangle=n\pi, n=0,\pm 1, \cdots$) 
\begin{equation}
\int\limits^{1/\ell}_{k_{min}} 
\left(\delta\rho_a^2\right)_k  d\ln k \simeq 
\frac{1}{4}  V_0^2\  e^{-\int\limits_{1/\ell}^{\infty} d\ln k
\frac{\delta\psi_k^2}{\psi_0^2}},  
\label{delrhof}
\end{equation}
and, finally, using $\langle\rho_a\rangle=\half V_0$ we obtain
\begin{equation}
\frac{\left(\delta\rho_a\right)_k}{\rho_a}\simeq  \frac{\delta\psi_k
}{\psi_0 }\  e^{-\half\int\limits_{k}^{k_S} d\ln k
\frac{\delta\psi_k^2}{\psi_0^2}},
\label{subt}
\end{equation}
where the upper limit on the right-hand-side of the previous equation has been
changed from $\infty$ to $k_S$ since we  take into account only fluctuations
produced during the dilaton-driven phase. The derivation of eq.(\ref{subt})
involves some subtleties which we will not discuss. Our result agrees with
the results of \cite{koflinde}.

 Because the standard deviation of
fluctuations,  $\sigma_k=\sqrt{\int\limits_{k}^{k_S} d\ln
k|\delta\psi_k|^2}$,  is much larger than the period of the axion potential,  
$\sigma_k \gg \psi_0$, energy fluctuations at large wavelength are exponentially
small, leading to the surprising conclusion that larger field fluctuations lead
to smaller energy fluctuations. Note that if the spectrum of perturbation is flat
as in  \cite{koflinde}, energy perturbations are only power-law suppressed. If,
as generally assumed in many cases $\delta\psi_k<\psi_0$ for all $k$, then
$\frac{\left(\delta\rho_a \right)_k}{\rho_a}\simeq  
\frac{\delta\psi_k }{\psi_0}$, and energy fluctuations actually grow as field fluctuations grow.

Finally, we have not considered any other axion production 
mechanisms, such as thermal production \cite{eu,axrev,batshel}, 
or the formations of strings, black holes, and other topological objects
\cite{lily,kotk}  which are likely to appear in our model because of the 
large field fluctuations and could result in additional and perhaps
dominant axion production  leading to a modification of our constraints. 
We hope to discuss these interesting alternatives in the near future.

The primordial axion spectrum is not thermal, and may consists of a 
fraction of relativistic axions even after their potential is generated. Our
understanding of the dynamics of the relativistic part of the spectrum is not
quite complete because after the axion potential is generated the problem
becomes an essentially non-linear problem. We believe that a better treatment
of the relativistic axions is interesting and should be done using numerical
 simulations and tools similar to those used in the theory of topological
defects. But we can nevertheless reach a few
conclusions. First, a necessary condition for a relativistic tail to exist after
the axion potential is fully developed is that $\omega_S(T\sim
\Lambda_{QCD})> m_a$, otherwise it can be shown that all modes
are non-relativistic. This condition leads to the
condition $z_S\laq\psi_0/\Lambda_{QCD}$. If $z_S\gaq\psi_0/\Lambda_{QCD}
\simeq 10^{10} (\psi_0/10^9 GeV) (100 MeV/\Lambda_{QCD})$, then all 
axions are massive.
Whether this $z_S$ range can be
consistent with condition (\ref{cond}) depends on $\psi_0$, $g_1$, $g_S$. For
$g_1\sim g_S$ the condition becomes $.1 g_1^5 M_p/\Lambda_{QCD}\laq
z_S\laq\psi_0/\Lambda_{QCD}$, requiring $.1 g_1^5 M_p \gaq \psi_0$, pushing
parameters into a relatively narrow region. Modes for which $\omega/\omega_S>
\Lambda_{QCD} z_S/\psi_0$ are relativistic at $T\sim\Lambda_{QCD}$.

The spectrum of axionic perturbations inside the horizon after the 
generation of the potential is quite complicated. A full treatment of these
perturbations is outside the scope of this paper, and may even result in the 
conclusion that for the particular case we are considering it is not allowed
to have any relativistic axions after the potential is generated.
 However, there is a range 
of frequencies for which we can nevertheless draw  definite conclusions. This
is the  upper end of the frequency range, for which the perturbation
is small $\delta\psi_k<\psi_0$,  and  relativistic $\omega>m_a$.
As the universe expands, kinetic energies redshift and more axions become
non-relativistic. We may evaluate  their energy density by calculating their
number just after the onset of the  potential, and, using number conservation,
calculate their energy density at  later times and in particular at
matter-radiation equality time. This was done in \cite{bh},
$\frac{d\Omega_a}{d\ln\omega}\simeq g_1^2\left(\frac{g_1}{g_S}\right)^2
\left(\frac{\omega}{\omega_S}\right)^{2-2\sqrt{3}}
\frac{\sqrt{m_a^2 + \omega^2}}{\omega_S}$.
 Evaluating the axion energy density at
matter-radiation equality, assuming that all particles have become
non-relativistic by then, we obtain
  \begin{equation}
\frac{d\Omega_a}{d\ln\omega}\simeq g_1^2\left(\frac{g_1}{g_S}\right)^2
\left(\frac{\Lambda_{QCD} z_S}{\psi_0}\right)^{3-2\sqrt{3}}
\frac{\Lambda_{QCD}}{T_{eq}}.
\end{equation}
Since $\frac{\Lambda_{QCD}}{T_{eq}}\sim 10^8$ we see that this region of
parameter space gives an uncomfortably large energy density, leading to
a seemingly favorable region of parameter space  
$z_S\gaq\psi_0/\Lambda_{QCD}$. However, since our
estimates are quite rough, and a small amount of entropy production during the
evolution of the universe may relax this condition we would not like at this
moment to completely rule out this interesting possibility. Note that in this
case even if $\psi_0\sim 10^9 GeV$ and the axion mass gets pushed towards its
upper limit $m_a\sim 10^{-2} eV$ axions can provide closure density. This is
important for their possible detection.

In general, for modified spectra, and other relic particles produced by
amplification of quantum fluctuations during the dilaton-driven phase, it may
well be that a fraction of relativistic particles remains at $t_{eq}$ and 
therefore it is possible that a single species provides simulatneously hot and
cold dark matter.

\section{Conclusions}

We have shown that relic axions are produced by amplification of quantum
fluctuations with a specific spectrum.  In some range of string cosmology
model parameters it is
predicted that most of the energy in our universe today is in the form of cold
dark matter axions,  with suppressed energy density fluctuations at large
wavelengths. Axions could provide closure density if their masses lie
in the  allowed range $10^{-6}eV\laq m_a\laq 10^{-2} eV$, depending on
parameters of string cosmology. The spectrum of primordial axions is not
thermal, and could contain a relativistic tail until quite late times.

\acknowledgments 
We thank L. Kofman and G. Veneziano for useful comments, 
and M. Schwartz for  helpful suggestions.
This work is supported in part by the  Israel
Science Foundation administered by the Israel Academy of Sciences and
Humanities.

\end{document}